\documentclass[runningheads]{llncs}
\usepackage[T1]{fontenc}
\usepackage{amsmath}
\usepackage[numbers,sort]{natbib}
\usepackage{forest}
\usepackage{graphicx}
\usepackage{svg}

\begin{document}
\title{Automating Categorization of Scientific Texts with In-Context Learning and Prompt-Chaining in Large Language Models}
\titlerunning{Automating Categorization of Scientific Texts using LLMs}
%
\author{Gautam Kishore Shahi\inst{1}\orcidID{0000-0001-6168-0132} \and \\
Oliver Hummel\inst{1}\orcidID{0009-0007-3826-9477} \
}
\authorrunning{Shahi \& Hummel}
%
\institute{Technische Hochschule Mannheim, Germany \\
\email{g.shahi@hs-mannheim.de}
}
\maketitle              
\begin{abstract}

The relentless expansion of scientific literature presents significant challenges for navigation and knowledge discovery. Within Research Information Retrieval, established tasks such as text summarization and classification remain crucial for enabling researchers and practitioners to effectively navigate this vast landscape, so that efforts have increasingly been focused on developing advanced research information systems. These systems aim not only to provide standard keyword-based search functionalities but also to incorporate capabilities for automatic content categorization within knowledge-intensive organizations across academia and industry. This study systematically evaluates the performance of off-the-shelf Large Language Models (LLMs) in analyzing scientific texts according to a given classification scheme. We utilized the hierarchical ORKG taxonomy as a classification framework, employing the FORC dataset as ground truth. We investigated the effectiveness of advanced prompt engineering strategies, namely In-Context Learning (ICL) and Prompt Chaining, and experimentally explored the influence of the LLMs' temperature hyperparameter on classification accuracy. Our experiments demonstrate that Prompt Chaining yields superior classification accuracy compared to pure ICL, particularly when applied to the nested structure of the ORKG taxonomy. LLMs with prompt chaining outperform the state-of-the-art models for domain (1st level) prediction and show even better performance for subject (2nd level) prediction compared to the older BERT model. However, LLMs are not yet able to perform well in classifying the topic (3rd level) of research areas based on this specific hierarchical taxonomy, as they only reach about 50\% accuracy even with prompt chaining.

\keywords{Large Language Models \and Field of Research Classification \and Prompt Engineering \and Prompt Chaining \and Scholarly data}
\end{abstract}
\section{Introduction}
\label{sec:introduction}

In recent years, the landscape of scholarly data has expanded significantly; however, handling and extracting meaningful information from this vast and growing amount of data remains a major challenge. Existing digital infrastructures require further development to ensure more user-friendly access to this valuable information \cite{hong2021challenges}.
The amount of scholarly publication is still consistently increasing worldwide; for example, around 3.3 million research articles have been published in 2023 alone \cite{nsf2023publication}. Over the past ten years, the total number of scientific publications worldwide has increased by 59\% per year Due to this enormous amount of data in numerous complex research areas, tagging a publication with a specific subject is challenging \cite{shahi2024enhancing}. Hence, classification of publications using artificial intelligence has been attracting increasing attention in recent years \cite{bornmann2021growth} as it seems to be a promising strategy for better organizing the body of knowledge in research and other knowledge intensive organizations. 
However, classifying the research area of scientific texts requires significant domain knowledge in various complex research fields. Hence, manual classification is challenging and time-consuming for librarians and limits the number of texts that can be classified manually by them \cite{zhang2023usage}. Moreover, due to the complex hierarchical classification schemes and their overwhelming variety, classification of scholarly  publications is also a challenging and unbeloved activity for researchers. 

Prominent examples of such classification schemes include the Open Research Knowledge Graph (ORKG) \cite{auer2019towards}, Microsoft Academic Graph \cite{wang2020microsoft}, the Semantic Scholar Academic Graph \cite{kinney2023semantic}, ACM computing classification system \cite{rous2012major}, Dewey Decimal Classification (DDC) \cite{scott1998dewey}, or the ACL Anthology \cite{bird2008acl}. Unfortunately, the coverage of these schemes is often subject-specific, for example, the mentioned ACM classification is merely limited to computer science topics. As another example, consider ORKG, which comprises three hierarchical categorization levels and currently has no in-depth classification for the top-level domain Arts and Humanities.\footnote{https://huggingface.co/spaces/rabuahmad/forcI-taxonomy/blob/main/taxonomy.json}

Additional challenges with the existing approaches arise in terms of the scalability of manual identification of research areas and are highlighted by the following examples. First, again consider ORKG, which was created only recently by a relatively small number of volunteers who were merely able to classify a few tens of thousands of publications with it so far. An automated classification engine would significantly help to increase its coverage more quickly. Similarly, Microsoft Academic Graph (MAG) used existing fields of study for scientific texts from the Microsoft Academic website; however, it has several limitations, such as incorrect tagging of subject areas due to immature automatic approaches or a non-optimal hierarchical structure \cite{herrmannova2016analysis}. Moreover, as Microsoft has discontinued its maintenance in 2021, there will be no updates for emerging research areas.
DDC, eventually, will always continue to struggle with newly emerging research topics and interdisciplinary fields \cite{wang2009extensive} as the maximum amount of represented categories is limited to one thousand by its design. 

Thus, within organizations such as universities, research institutes, or even large companies where numerous researchers and other knowledge creators are working in multiple diverse domains, categorizing texts still requires considerable manual effort, making it challenging to deal with the huge volume of created texts in order to better organize and access the knowledge contained. Consider even relatively simple demands, such as identifying domain experts as an example that is not possible without an assignment of publications to subject areas. Consequently, there is a serious demand for automated subject tagging systems to help efficiently managing the steadily increasing volume of scientific texts and general knowledge contained in institutional repositories and comprehensive digital archives.

With the growth in generative artificial intelligence (GAI), especially, Large Language Models (LLMs) \cite{zhao2023survey}, a new opportunity to automate tagging of scholarly publication has finally become tangible. LLMs are Artificial Intelligence (AI) systems that are specialized in generating human-like text for different tasks such as summarization, translation, content creation, and even coding. Consequently, LLMs have already been applied for several similar use cases, such as analyzing scientific documents \cite{giglou2024llms4synthesis},
writing scientific reviews \cite{mahapatra2024artificial}, or information extraction and summarization \cite{pertsas2024annotated}.
LLMs can be configured by setting parameters such as their so-called temperature, which controls the degree of ``creativity'' (i.e. randomness) in an LLM's answer. LLMs are applied to their respective tasks by using so-called prompts, which are essentially textual commands describing the desired task at hand. The proper engineering of these prompts plays an important role in achieving the desired results with a model invocation \cite{gao2023prompt}.

\subsection{Research Goals}
\label{rqs}
With this study, we aim to better understand the benefits and quality currently achievable when using ``off-the-shelf'' LLMs without fine-tuning for the classification of scientific texts from scholarly publications and hence propose the following research questions. 

\noindent \textbf{RQ1:} 
How can LLMs be effectively used to perform accurate tagging of research areas based on existing taxonomies?

\noindent To address the RQ1, we employed the OKRG taxonomy\footnote{\url{https://orkg.org/fields}}
 as a hierarchical classification scheme, alongside the recently introduced Field of Research Classification (FoRC) Shared Task dataset \cite{abu2024forc} as the ground truth. The FoRC dataset was constructed by aggregating manuscripts from ORKG \cite{jaradeh2019open} and arXiv, which were subsequently annotated by human experts across three hierarchical levels defined by the OKRG taxonomy.

We evaluated several contemporary open-source large language models (LLMs) on their ability to predict the research area of scientific publications based solely on their titles and abstracts, employing prompt-engineering techniques across different temperature settings. The results indicate that LLMs achieve the best performance at a temperature of 0.8, demonstrating strong alignment with human-assigned research area classifications.



\noindent \textbf{RQ2:} How does the choice of the prompting method influence the ability of Large Language Models to ``understand'' the hierarchical structure of ORKG taxonomy while annotating scientific texts?

\noindent To address the second research question (RQ2), we experimented with various prompt engineering strategies to assess whether LLMs are capable of capturing the hierarchical structure of the taxonomy. Specifically, we applied in-context learning and prompt chaining techniques. In-context learning was explored through zero-shot, one-shot, and few-shot settings. The zero-shot prompt included only the task description and the ORKG taxonomy; the one-shot prompt added a single classification example; and the few-shot prompt provided several examples of research papers along with their assigned research areas.

In prompt chaining, the multi-level classification process defined by the ORKG taxonomy is decomposed into a series of structured steps. Instead of instructing the LLM to directly predict the most specific (lower-level) research area, an iterative top-down approach is applied. Each text is classified progressively at different levels of the hierarchy. Specifically, we first provided the LLM with the four top-level domains (1st level) from the taxonomy and asked it to predict the appropriate domain for a given text. Once the domain was identified, the next prompt included the corresponding list of subjects within that domain, prompting the LLM to select the most suitable subject (2nd level). Finally, after determining the subject, the LLM was asked to choose the best matching topic (3rd level) within that subject.

This iterative procedure effectively reduces the overall complexity of navigating a nested taxonomy, as the model operates within a constrained set of options at each step. Such hierarchical prompt chaining is designed to enhance classification accuracy, minimize ambiguity, and better leverage the LLM’s reasoning capabilities in a structured manner. The proposed approach was compared against zero-shot, one-shot, and few-shot prompt engineering strategies under different temperature settings. Detailed prompt formulations are provided in Section \ref{sec:2.1}.

Model performance was primarily evaluated using accuracy. However, rather than relying solely on direct accuracy scores, we further incorporated refined evaluation metrics—including exact match, string distance, and embedding distance—as described in Section \ref{sec:metric}. These additional metrics capture partially correct predictions, accounting for semantic and lexical proximity between predicted and ground-truth labels



The present study is an extended version of our previous work \cite{iceis25}, which primarily focused on exploring the use of LLMs for predicting research areas from scientific texts, as outlined in 1st Research Question 1 (RQ1). In contrast, the current study expands upon that foundation by addressing 2nd Research Question 2 (RQ2), aiming to provide a deeper understanding of the capabilities of LLMs in capturing and predicting the hierarchical structure of research areas. Furthermore, this extended work offers a more comprehensive analysis of the impact of prompt engineering techniques and the alignment between LLM-generated predictions and human annotations on overall prediction quality. Hence, the key contributions of this paper are as follows; it presents:

\begin{itemize}
   \item A classification approach for identifying the research areas of scientific texts, introducing a novel application of LLMs in research domain classification.
\item An approach for hierarchical prediction of research areas across three levels of the ORKG taxonomy — domain (1st level), subject (2nd level), and topic (3rd level).
\item A comprehensive evaluation of prompt engineering techniques, including in-context learning (zero-shot, one-shot, and few-shot) and prompt chaining, combined with parameter tuning for improving the classification performance of LLMs on scientific texts.
\end{itemize}

In the remainder of this paper, we discuss the state of the art in section~\ref{sec:related} and the proposed approach itself in section~\ref{sec:method}. After that, we discuss the implementation of our enhancements of the proposed approach in section~\ref{sec:implementation} and present results in section~\ref{sec:result}. Finally, we conclude our work and discuss future work in section~\ref{sec:future}.


\section{Related Work}
\label{sec:related}


Document classification is one of the primary tasks for classifying scholarly research that is usually either performed by librarians or by subject experts, where both groups are faced with individual challenges: while the former are usually not subject experts, the latter are normally not trained for using document classification schemes as mentioned above. 

Until now, the automated classification of scientific articles into their respective research fields is -- despite decades of research -- still rather an emerging discipline \cite{desale2014research} than a proven practice that can be applied in libraries, universities, or the knowledge management of large corporations. In previous works, multiple approaches have been applied for this challenge, for instance, a supervised machine-learning approach used for assigning DDC identifiers to documents collected from the Library of Congress achieved an accuracy of more than 90\% \cite{wang2009extensive} based on a significant previous training effort. Golub et al. used six different machine learning algorithms to classify documents from the Swedish library, where a Support Vector Machine gave the best results with 81.9\% accuracy \cite{golub2020automatic}. Golub et al. used BERT, an early transformer model, for the identification of the research area on previously annotated data \cite{jiang2020improving}. However, up until today, the automation of document classifications has mainly been carried out as a supervised learning task that requires specific training data and a thorough training period. 

Moreover, additional challenges, such as the deep nesting of many classification taxonomies and data sparseness in certain classes, also need to be taken into account when implementing classification with ``traditional'' supervised learning. In general, this is especially challenging due to the need for a significant amount of labeled training data that is still hard to find today \cite{kalyan2023survey}. However, as the recent generation of Large Language Models is pre-trained on extensive bodies of text data, they are supposed to be more proficient in generating high-quality text classifications without additional training or finetuning.

With the recent advancement of LLMs, these models have already been tested for several generic tasks in analyzing scholarly writing and have provided promising results. In one recent study, ChatGPT has been used for automated classification of undergraduate courses in an e-learning system and yielded an overall precision of 0.73 on the one hand; however, it also delivered additional hallucinated results in 48\% of the cases \cite{young2024chatgpt}. In another study, the authors used LLMs for the automatic evaluation of scientific texts (in German language) written by students to assign a grade \cite{abburi2023generative}. Pal et al. proposed an approach for using ChatGPT to develop an algorithm (computer program) for ensuring plagiarism-free scientific writing \cite{pal2024ai}. Previously, authors built a data set to detect machine-generated scientific papers and compared results with other benchmark models \cite{mosca2023distinguishing}. 

However, to our knowledge, so far, LLMs have not been tested for the identification of research areas of scientific texts, and hence our work provides a novel insight into the current performance of off-the-shelf LLMs in this area. To the best of our knowledge, a recent publication \cite{shahi2024enhancing}, where authors proposed enhancing a search engine for scientific documents with research areas, is the first published work in this specific direction. There, we have illustrated the practical usability of such subject tagging, e.g., for searches of domain experts. 

In summary, the automated classification of scientific articles into their respective research fields is still an emerging area of research \cite{desale2014research}. So far, academic institutions and other knowledge-intensive organizations have largely relied on manual methods for gathering and tagging scholarly publications. As previous research on automating this tedious task has been, to the best of our knowledge, constrained to relatively primitive supervised learning approaches, using LLMs for this task can lead to a significant advancement of the state of the art.


\section{Foundations}
\label{sec:2}
Our contribution relies on solid foundations from the areas of Generative AI and bibliographic classifications schemes, which we will introduce in the following.

\subsection{Generative Artificial Intelligence}
\label{sec:2.1}
The key aspect of Generative Artificial Intelligence (GAI) that separates it from other forms of artificial intelligence (AI), is that it is not primarily dedicated to analyzing (numerical) data or acting based on such data like ``traditional'' AI (i.e. machine learning approaches) that has mainly been used for this purpose in the past. Instead, GAI focuses on creating new content by using the very data it was trained upon \cite{hacker2023regulating,murphy2022probabilistic}. \textit{The term GAI thus refers to computational approaches which are capable of producing apparently new, meaningful content such as text, images, or even audio and video} \cite{feuerriegel2024generative}.

Modern GAI for texts utilizes so-called Large Language Models (LLMs) that have been trained on massive datasets to acquire different capabilities such as content generation or text summarization by learning the statistical relationships of words or word components \cite{wang2024gpt}. Modern LLMs are developed based on the so-called transformer architecture \cite{vaswani2017attention} and trained on extensive corpora collected from public sources such as Web Crawls or digital platforms such as Wikipedia or GitHub using self- and human-supervised learning, enabling them to capture complex language patterns and contextual relationships \cite{perelkiewicz2024review}. Hence, LLMs can also be used for other quite diverse applications in natural language processing such as text summarization or data annotation. The well-known ChatGPT, launched by OpenAI, based on the Generative Pre-trained Transformer (GPT) architecture \cite{nah2023activity} is one such GAI that has been trained on a huge body (i.e., a significant part of the public WWW) of text.

LLMs are AI models trained on a larger corpus, with billions of parameters, capable of deep reasoning, creativity, and complex tasks—but they require a significant amount of computing power. SLMs (Small Language Models) are lightweight, faster, and cheaper to run, making them ideal for domain-specific or straightforward tasks on smaller devices. Often, SLMs are trained on domain-specific datasets rather than a huge Web corpus.



\subsection{Classification Taxonomy}
\label{sec:2.2}

Taxonomies provide structured frameworks to organize and classify scientific text, helping researchers and practitioners navigate large and complex collections of documents. Over the past decades, numerous taxonomic schemes have been developed to organize and classify the ever-growing volume of scientific and domain-specific literature. Notable examples include the ACM Computing Classification System \cite{rous2012major}, the Dewey Decimal Classification (DDC) \cite{scott1998dewey}, and the ACL Anthology \cite{bird2008acl}. Despite—or perhaps because of—the variety of existing classification schemes, manual subject tagging remains both challenging and time-consuming. In this study, we used ORKG taxonomy, which was created for development of ORKG knowledge graph \cite{jiang2020improving}.

\begin{figure}[ht]
    \centering
\resizebox{\textwidth}{!}{%
\begin{forest}
for tree={
    grow=east,
    draw,
    rounded corners,
    align=center,
    edge={->},
    parent anchor=east,
    child anchor=west,
    l sep+=15pt,
    s sep+=10pt,
    anchor=west,
}
[ORKG
    [\textbf{Humanities and Social Sciences}
        [\textit{Humanities}
            [Developmental and Educational Psychology]
        ]
        [{...}] 
        [\textit{Social and Behavioural Science}
            [Asian Studies]
        ]
    ]
    [\textbf{Life Sciences}
        [\textit{Biology}
            [Biochemistry]
        ]
        [{...}] 
        [\textit{Medicine}
            [Virology]
        ]
    ]
    [\textbf{Natural Sciences}
        [\textit{Geosciences}
            [{Hydrogeology, Hydrology, Limnology, Urban Water Management, ...}]
        ]
        [{...}] 
        [\textit{Thermal Engineering/Process Engineering}
            [Biological Process Engineering]
        ]
    ]
    [\textbf{Engineering Sciences}
        [\textit{Construction Engineering and Architecture}
            [{Urbanism, Spatial Planning, ...}]
        ]
        [{...}] 
        [\textit{Materials Science and Engineering}
            [Biomaterials]
        ]
    ]
    [\textbf{Arts and Humanities}
        [\textit{Philosophy}
            [Esthetics]
        ]
        [\textit{History}
            [European]
        ]
    ]
]
\end{forest}
}
\caption{Hierarchical structure of domains, subjects, and topics within ORKG taxonomy.}
\label{fig:domain-tree}
\end{figure}
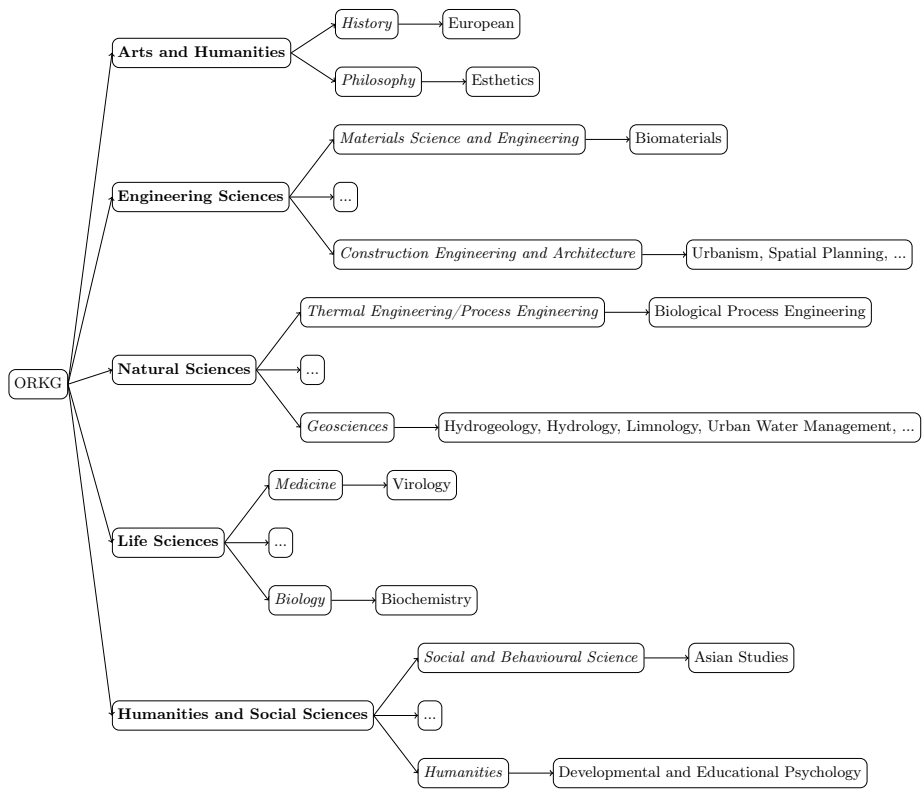

The ORKG taxonomy provides a structured framework for the systematic classification and exploration of research domains. This taxonomy is organized into five primary domains: \textit{Arts and Humanities, Engineering, Life Sciences, Physical Sciences and Mathematics, and Social and Behavioral Sciences}. Each of these domains is hierarchically structured into two additional levels: subdomains and subjects. At the first sub-level, each primary domain is subdivided into subjects, which are further refined into specialized topics. For instance, within the \textit{Physical Sciences and Mathematics} domain, the \textit{Computer Science} subdomain includes \textit{Artificial Intelligence} as a subject. Some examples of domains, subjects, and topics are given in Table~\ref{fig:domain-tree}. The root node displays the ORKG taxonomy; the first child represents \textit{domains}, the second child represents \textit{subjects}, and the last represents \textit{topics}. In this study, we aim to classify scientific texts according to a taxonomy structure. For instance, a scientific text is classified as belonging to the domain of Life Science, followed by the subject \textit{Medicine} and the topic \textit{Virology}.


There are several limitations in the ORKG taxonomy. First, it covers only 14 subjects and does not fully span all four domains. Moreover, some subject names are multivalent—for example, Thermal Engineering/Process Engineering—which could be mapped as equivalent classes in an ontology, but remain challenging for LLMs to predict accurately, even with different evaluation techniques. Similarly, many subjects are multi-label, such as Urbanism, Spatial Planning, Transportation and Infrastructure Planning, Landscape Planning or Hydrogeology, Hydrology, Limnology, Urban Water Management, Water Chemistry, Integrated Water Resources Management, which further complicates correct topic prediction for LLMs.

\subsection{Dataset}
\label{sec:2.3}

For our evaluations, we used scientific texts collected by the FORC shared task \cite{abu2024forc}, which is classified using the ORKG taxonomy. FORC consists of scientific texts, mainly research papers with DOI, research area, abstract, title, and author information. The FORC initiative compiled scientific texts from open-source resources such as ORKG (CC0 1.0 Universal) and arXiv (CC0 1.0), whereas scientific text with non-English titles or abstracts were excluded. Each scientific text has been assigned a field of research based on the ORKG taxonomy.\footnote{https://orkg.org/fields} 

In FORC dataset, the issue of mulitvanet subjects and topics are there but also there is no scientific text for class Arts and Humanities.

\section{Approach}
\label{sec:method}
\begin{figure*}[!htbp]
  \centering
\includegraphics[width=0.99\linewidth]{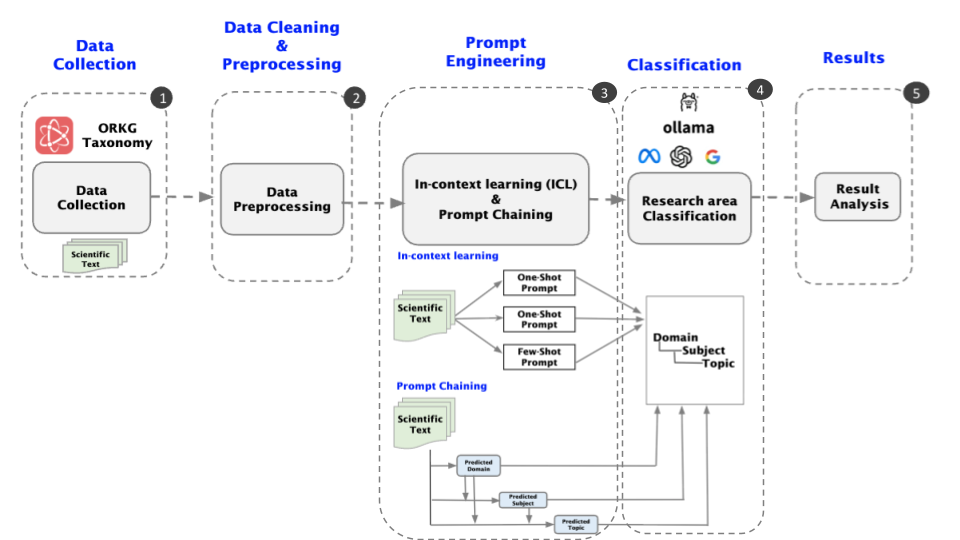}
  \caption{Methodology used in the identification of research area and upper part shows hierarchical identification of research areas using In-context learning and Prompt Chaining)}
  \label{fig:method}
 \end{figure*}

To address the proposed research questions, we designed a rigorous five-step methodological procedure summarized in Figure~\ref{fig:method} and detailed in the following subsections: data collection; data cleaning and preprocessing; prompt engineering; research area classification; and results analysis.

\textbf{Data Collection}
The initial step involved the selection and collection of a suitable dataset. For this study, we analyzed scientific texts sourced from the FORC dataset. A comprehensive description of the dataset, including its structure and composition, is provided in Section~\ref{sec:dataset}.

\textbf{Data Preprocessing}
The second step focused on preparing the data for ingestion. This involved extracting the paper title and abstract from the source data and implementing a cleaning process to remove extraneous formatting information and other unwanted elements prior to feeding the texts to the Large Language Models (LLMs).

\textbf{Prompt Engineering} The third step, prompt engineering, explored various prompt settings and configurations. Specifically, we investigated the efficacy of different prompting strategies, including In-Context Learning (ICL) and Prompt Chaining, to optimize the LLM performance.

In this study, we employ prompt engineering in two settings, i.e with ICL and Prompt Chaining. Both prompt engineering approaches aim to better predict the hierarchical classification of research areas for the ORKG taxonomy.

In ICL, we use zero-shot, one-shot, and few-shot prompting for invoking LLMs and ask them to predict the domain, subject, and topics of given scientific text. A detailed description of different prompts and examples is provided in Section~\ref{sec:prompt}.


The research area classification is designed as a hierarchical process, where scientific texts are analyzed to predict research areas across three distinct levels: domain, subject, and topic. This multi-level approach enables the systematic categorization of research texts, progressing from broad disciplinary domains to more specific subjects and, ultimately, to precise research topics, thereby capturing the full hierarchy of research areas. As illustrated in Figure~\ref{fig:method} (upper part), the process begins by providing a list of domains from the OKRG dataset to the language model, which predicts the most relevant domain for a given research text. Once the domain is identified, a corresponding list of potential subjects within that domain is supplied to the model to predict the most suitable subject. Finally, after determining both the domain and subject, a list of relevant topics is provided, and the model predicts the most specific topic associated with the same text. The outcome of this stepwise process is a hierarchical classification of each research text into its domain, subject, and topic levels.

The hierarchical identification of research is aligned with taxonomy explained in Figure~\ref{fig:method}, i.e, we provided scientific text to LLM and if domain is identified as \emph{Life Sciences} then a list of subject is provided as \emph{[Biology, Medicine, ....]} then ask LLM to predict subject followed by provided lists of topics as \emph{[Biochemistry, Virology, ....]} and ask to predict topic of scientific text.

\textbf{Research Area Classification}
The fourth and most crucial step involved applying the LLMs to the scientific texts to perform research area classification and subsequently analyze the generated predictions. A central aim of this study is the prediction of all three hierarchical levels of the ORKG taxonomy: domain, subject, and topic.

\textbf{Results Analysis}
The final step involved a detailed analysis of the classification outcomes. The entire methodology is further elaborated in the following subsections.




\subsection{Dataset}
\label{sec:dataset}

Overall, FORC provides a collection of 50,441 scientific texts, each categorized using a taxonomy of 123 Fields of Research (FoR). 
For each scientific text, we used the DOI as a unique identifier while title and abstract were fed to our classification model to predict the research area. A detailed description of the used metadata is provided in Table~\ref{dataset}.

\begin{table*}[!htb]
\centering
\caption{Description of field used from FORC dataset \cite{iceis25}}
\begin{tabular}{|p{2.5cm}|p{5cm}|p{4.5cm}|}
\hline
\textbf{Field} & \textbf{Description} & \textbf{Example} \\
\hline

DOI & A DOI (Digital Object Identifier) is a standardized unique number given to research papers, hence we were able to use it as a unique identifier for each paper & 10.1145/2736277.2741136 \\ \hline
Title & Title describes the title of the paper &  Crowd Fraud Detection in Internet Advertising \\ \hline
Abstract & Abstract of the paper describing a summary of the content & ``the rise of crowdsourcing brings new types of malpractices in internet advertising. one can easily hire web workers through malicious crowdsourcing platforms to attack other advertisers...'' \\ \hline
Research area & research area defined based on the ORKG taxonomy, and it is a dependent variable for our prediction model & Engineering \\ \hline 
 
\end{tabular}
\label{dataset}

\end{table*}

\subsection{Data Cleaning and Preprocessing}
This section describes the steps involved in the data preprocessing, such as removing unwanted artifacts left over from text parsing. In principal, LLMs should be able to ``circumvent'' such artifacts, however, LLMs work as a black-box, hence we cannot be sure whether LLMs might be confused by such artifacts \cite{liu2024language}.
Moreover, we excluded scientific texts that did not contain an abstract, as relying solely on titles can make it challenging to accurately predict the research area. To address the issue of sparse topic and subject distributions, we identified subjects with low frequencies and removed those with fewer than 100 occurrences. 
We provided cleaned data for all models to maintain fairness. After collecting the dataset, we also removed unwanted information, such as URLs mentioned in the text, special characters in abstracts, and authors names of the publication. After preprocessing, a total of \emph{24,911} scientific text were retained for analysis. The final dataset encompasses four domains, 14 subjects, and 40 topics in total.


\subsection{Prompt Engineering}
\label{sec:prompt}
The so-called prompt engineering is an important factor that can significantly influence the performance of an LLM for a given task. Prompting strategies in LLMs include writing instructions for the models that are intended to guide responses effectively. Common techniques include providing context, step-by-step instructions, and examples to improve accuracy and relevance. A more detailed and comprehensive description of prompting strategies can, e.g., be found in \cite{al2025evaluation}. In our experiments, we used the following two prompting strategies for evaluating the LLM and provide some examples from our context in Table~\ref{prompt}.

\begin{itemize}
     \item \textbf{Zero-shot} In this approach, we ask the LLMs to annotate the research area without providing any description or examples, which employs the most simple and straightforward approach. Zero-shot prompts are also known as vanilla prompts, that do not take any prior knowledge or specific training on a given task into account. It uses the pre-trained general ``intelligence'' of an LLM to obtain the research area for a scientific text. 
 \item  \textbf{One-shot} In this case, learning is done based on context, hence the term In-Context Learning, where the model takes some description and an example for the research area as defined by the ORKG taxonomy to better understand the task at hand. The model takes this input and provides answers based on the given information in conjunction with its previously learned general knowledge.
  \item  \textbf{Few-shot} In this case, learning is done based on an extended context where the model takes a few examples of different classes for the research area as defined by the ORKG taxonomy to better understand the task at hand. The model takes these inputs and provides answers based on the given information in conjunction with its previously learned general knowledge.
\end{itemize}

\begin{table*}[h!]
\centering
\caption{Prompting strategies for determining research area from scientific texts}
\begin{tabular}{|p{4.1cm}|p{4.1cm}|p{4.1cm}|}
\hline
\multicolumn{3}{|c|}{\textbf{In-context learning} } \\
\hline
\textbf{Zero-shot Prompt} & \textbf{One-shot Prompt} & \textbf{Few-shot Prompt} \\
\hline
Suppose you are a data annotator who finds the research area of scientific texts using ORKG taxonomy. & Suppose you are a data annotator who finds the research area of scientific texts. & Suppose you are a data annotator who finds the research area of scientific texts. \\

& You are provided with one example of scientific text with its research area. Your task is to read texts and determine which research area from the list best represents the content of the scientific texts. Here is the hierarchy for each research: \textit{taxonomy of research field extracted from ORKG} & 

You are provided with few examples of scientific texts with its research area. Your task is to read texts and determine which research area from the list best represents the content of the scientific texts. Here is the hierarchy for each research: taxonomy of research field extracted from ORKG\footnote{https://orkg.org/fields}  \\

Scientific text to annotate is: \textit{factors influencing the behavioral intention to adopt a technological innovation from a developing country context: the case of mobile augmented reality games} & Scientific text to annotate is \textit{comparative analysis of algorithms for identifying amplifications and deletions in array cgh data} &  

Scientific text to annotate is: \textit{comparative analysis of algorithms for identifying amplifications and deletions in array cgh data} \\
Assign a research area to the given scientific texts and provide it as output & Assign a research area from the given taxonomy above and provide it as output & Assign a research area from the given taxonomy above and provide it as output \\
\hline
\end{tabular}
\label{prompt}
\end{table*}

\begin{table*}[!htb]
\centering
\caption{Prompting strategy with Prompt Chaining for determining research areas from scientific texts on the three different levels of ORKG.}
\begin{tabular}{|p{2.2cm}|p{10cm}|}
\hline
\multicolumn{2}{|c|}{\textbf{Prompt Chaining} } \\
\hline
\textbf{Hierarchal Level} & \textbf{Prompts} \\
\hline
Domain & Suppose you are a data annotator who finds the research area from a scientific text. \\
& There are three hierarchical levels of annotation: domain, subject, and topic. \\
& Given a title and abstract from a scientific paper, extract only the domain. \\
& The list of valid domains is: [domain\_list]. \\
& Given this paper: (text of a research paper), return only one of the domain names exactly as written, without providing any explanation. \\

Subject & Once you have identified the \textit{Domain} of the scientific text in the above step, your task as a data annotator is to identify the subject of the text.  \\
& From the following list of subjects under this domain: [subject\_list] (note that each domain has its own list of subjects), \\
& given this paper: (text of a research paper), identify and return only one subject from the list above, without providing any explanation. \\

Topic & Once you have identified a \textit{Domain} and respective \textit{Subject} of the scientific text, your task is to identify the topic. \\
& From the following list of topics under this subject: [topic\_list] (note that each subject has its own list of topics), \\
& given this paper: (text of a research paper), identify and return only one topic from the list above, without providing any explanation. \\
\hline
\end{tabular}
\label{prompt}
\end{table*}

\subsubsection{Parameter Optimization}

Moreover, we have employed the LLMs with varying temperatures, which adjusts the randomness of the responses given by an LLM. Lower temperatures give more focused and deterministic results, while higher temperatures generate more diverse and ``creative'' answers. The value of the parameter starts from 0 and can be increased extensively; however, we limited it to the range of [0,1] as temperatures above 1 typically result in a very high degree of randomness and neither provide good coherence nor good reproducibility.

As different LLMs are trained with different objectives and with different training datasets, we assume that this is likely affecting their strengths in producing helpful results, i.e. categorizations, in our context. Hence, we not only used a set of different temperatures and prompts, but also different LLMs, such as Llama and Gemma (more details follow in section \ref{llm}), to let them identify the research area of scientific texts from the test data set.

\subsection{Classification of Research Area}

\subsubsection{Evaluated LLMs.}
\label{llm}

The next step is to using LLMs for classification of research areas of given scientific text. For that, we used a combination of title and abstract from the text of a presentation as input to the LLMs and asked the LLMs to predict the research area using different prompts settings. We employed four contemporary LLMs (cf. Table \ref{tab:llm}) with a small and medium amount of parameters between 3.82b and 70.4b to classify the research areas of the selected texts. Each LLM was obtained from and executed with Ollama\footnote{https://ollama.com/library}.
To assess the performance of LLMs, we compared their results with those of previous traditional Bidirectional Encoder Representations from Transformers (BERT) models \cite{devlin2018bert}. A detailed explanation of the experimental setup is provided in Section \ref{sec:implementation}, while the results are discussed in Section \ref{sec:result}. The models we have used are as follows:

\begin{itemize}
\item \textbf{Gemma} We used the recent version Gemma 2 \cite{team2024gemma}, which is a family of lightweight, state-of-the-art open-source models that are advertised as high-performing and efficient models by Google. They are currently available in two sizes; we have used Gemma 2 with 27 million parameters. Gemma was trained on web documents and using mathematics, outperforming other models in 11 of 18 text-based tasks in terms of efficiency \cite{team2024gemma}. 

\item \textbf{Llama} We used the latest version, which was Llama 3.1 at the time of writing. Llama is developed and released by Meta \cite{touvron2023Llama}; there are currently three versions of Llama with different sizes of 8b, 70b, and 405b parameters; we have used Llama 3.1 with 70b parameters. Llama is trained on publicly available data without resorting to proprietary datasets. For the training, different data sources, such as CommonCrawl and GitHub, were used.

\item \textbf{Mistral Nemo} is the latest LLM developed jointly by Mistral AI and NVIDIA AI with 12B parameters and a context window of up to 128k tokens. Mistral Nemo outperformed the prior Mistral model LLama 3 and Gemma 2 in terms of efficiency and effectiveness despite having fewer parameters. \footnote{\url{https://mistral.ai/news/mistral-nemo/}}

\item \textbf{Phi} is a family of lightweight, open large language models developed by Microsoft that are designed to be efficient and accessible. The Phi-3 family includes models with 3 billion (3B) and 14 billion (14B) parameters, classified as "Mini" and "Medium" respectively. Phi-3 outperforms\footnote{https://azure.microsoft.com/de-de/blog/new-models-added-to-the-phi-3-family-available-on-microsoft-azure/} Gemini 1.0 Pro, and the model is trained on high-quality educational data, newly created synthetic, “textbook-like” data, which should make it especially suitable for use for classification tasks of the scientific domain.
\end{itemize}

\begin{table*}[!htbp]
\centering
\caption{A short overview of different LLM models used in the study \cite{iceis25}}
\begin{tabular}{|p{2.2cm}|p{5cm}|p{3cm}|p{1.8cm}|}
\hline
\textbf{Model)} & \textbf{Description}  & \textbf{No. of Parameters}  & \textbf{Release Date}\\
\hline

Gemma 2\footnote{https://blog.google/technology/developers/google-gemma-2/} & Gemma2 is a lightweight, state-of-the-art open-source model &  parameters-27.2B \& quantization-Q4\_0 & June 2024  \\ \hline
Llama 3.1\footnote{https://ai.meta.com/blog/meta-llama-3-1/} & Llama 3.1 70B is a multilingual model that has a significantly longer context length of 128K, state-of-the-art tool use, and overall stronger reasoning capabilities  & parameters-70.4B and quantization-Q4\_0  & July 2024  \\ \hline

Mistral Nemo\footnote{https://mistral.ai/news/mistral-nemo/} & Mistral NeMo offers a large context window of up to 128k tokens. Its reasoning, world knowledge, and coding accuracy are state-of-the-art in its size category & parameters-12.2B \& quantization-Q4\_0 & July 2024 \\ \hline
Phi\footnote{https://huggingface.co/microsoft/Phi-3.5-mini-instruct} &  Phi 3.5 is a lightweight, state-of-the-art open model built upon synthetic datasets & parameters-3.82B \& quantization-Q4\_0 & August 2024 \\ \hline

\end{tabular}
\label{tab:llm}

\end{table*}

\section{\uppercase{Experiments}}
\label{sec:implementation}

For executing the experimental setup, we have been using the Ollama frameworks \cite{ollama}, an open source application that allows the easy execution of LLMs on local hardware. Ollama offers a straightforward way to run the model locally, featuring a simple command-line interface that directly interacts with the LLMs and facilitates easy installation and implementation. Ollama allows downloading models with a given number of parameters. The computational infrastructure consists of an in-house server equipped with four NVIDIA RTX A6000 GPUs, each with 48 GB of VRAM, a total amount of 512 GB of system memory, and 8 TB of storage, ensuring sufficient resources for efficient model execution and experimentation.

We developed a Python program utilizing Ollama and LangChain\footnote{https://python.langchain.com/docs/integrations/llms}, an open-source framework for building large language model (LLM) applications, for prompting the language models with our test data set for research area annotation. The generated results were systematically stored for each evaluation round with different parameters. To evaluate different configurations, we stimulated the models with various combinations of temperature settings and prompts. In the zero-shot setting, only the temperature and task description were provided, whereas the few-shot setting involved different prompt combinations to refine the outputs. Corresponding examples are presented in Table \ref{prompt}.

\subsection{Baseline models}
Given our collected corpus of scientific texts, we chose the following two classification models as (pre LLM) state-of-the-art methods for result comparison: \\
\textbf{BERT} \cite{devlin2018bert} is a widely used pre-trained model for text classification. The model has been applied to various classification tasks and evaluated across multiple domains, including the classification of text related to COVID-19 \cite{shahifakecovid}. BERT utilizes a bidirectional transformer mechanism, allowing it to capture contextual relationships in text more effectively than traditional models. It has demonstrated state-of-the-art performance in numerous natural language processing (NLP) benchmarks, making it a suiting candidate for research area classification.  \\
\textbf{BiLSTM} (Bidirectional Long Short-Term Memory) \cite{huang2015bidirectional} is a recurrent neural network (RNN) designed for text classification, capturing input flows in both forward and backward directions. It has been successfully applied to various NLP tasks, including the classification of scientific texts \cite{enamoto2022multi}. BiLSTM enhances sequential data processing by preserving long-range dependencies, reducing the vanishing gradient problem, and improving contextual understanding. Its ability to capture bidirectional dependencies makes it effective in tasks requiring nuanced text comprehension.

For the utilization of the baseline models, we also developed a Python program and retrieved the pre-trained models from Hugging Face.\footnote{\url{https://huggingface.co/}} Both BERT and BiLSTM were implemented using the models available on this platform.

\subsection{Evaluation Metrics}
\label{sec:metric}

LLMs are well known for somewhat ``creative'' answers and not complying with all specifications under all circumstances. Moreover, even with good specifications, LLMs might produce some ``near misses'' with their results when, for instance, a research area like ``Computer Science \& Engineering'' might be predicted as just ``Computer Science''. To overcome these challenges, we used different approaches for determining the actual matches as described below. Based on that, we used accuracy for calculating the overall performance of different LLMs as recommended e.g. by Chang et al. \cite{chang2024survey}.

\begin{itemize}
    \item \textbf{Exact Match (EM)}  In exact match, we compare the output of the model directly with the ground truth data. It is a binary measure for comparing the result; if it matches, then 1, else 0. For instance, if the LLM returns ``Social'' and the actual research area is ``Social Science'', then the EM returns 0.
    
    \item \textbf{String Distance (SD)} In the string distance, we calculate the closeness of predicted research areas and actual research areas by computing Levenshtein distance \cite{devatine2024assessing}. To do so, we utilize the normalized Levenshtein distance of the expected result and the actual result produced by LLMs. If the value is greater than 0.7, we counted it as a match. The benefits of these measures are that they capture partially correct answers, where the phrasing may differ slightly, but still is close to the expected result. The normalized Levenshtein distance is calculated using the formula as defined in equations~\ref{eq:ld1} and ~\ref{eq:ld2}.

\begin{equation} \label{eq:ld1}
\text{String Distance}(s_1, s_2) = 1 - \frac{lev(s_1, s_2)}{\max(|s_1|, |s_2|)}
\end{equation}

\begin{equation} \label{eq:ld2}
\begin{aligned}
&s_1, s_2 &&\text{are the two strings being compared},\\
&lev(s_1, s_2) &&\text{is the Levenshtein distance between } s_1 \text{ and } s_2,\\
&|s_1|, |s_2| &&\text{are the lengths of the strings},\\
&\max(|s_1|, |s_2|) &&\text{ensures normalization between 0 and 1}.
\end{aligned}
\end{equation}


    \item \textbf{Embedding Distance (ED)}  In this measure, we compare the semantic similarity between the result obtained from the model and the actual research area using vector embeddings. For computing the similarity, we have used BERT sentence embedding \cite{reimers-2019-sentence-bert}, where we transform the input into vectors and calculate the similarity score as defined in equation~\ref{eq:ed1}. If the value is greater than 0.7, then it is counted as a match; otherwise, it is not as defined in equation~\ref{eq:ed2}. The benefit of this measure is that different words that convey the same meaning can be considered as a match.

    Let $\mathbf{v}_\text{model}$ and $\mathbf{v}_\text{actual}$ be the sentence embeddings obtained from BERT for the model's output and the actual research area, respectively \cite{reimers-2019-sentence-bert}. Then the semantic similarity score $ED$ can be computed using the cosine similarity:

\begin{equation} \label{eq:ed1}
ED = \cos(\mathbf{v}_\text{model}, \mathbf{v}_\text{actual}) 
  = \frac{\mathbf{v}_\text{model} \cdot \mathbf{v}_\text{actual}}{\|\mathbf{v}_\text{model}\| \, \|\mathbf{v}_\text{actual}\|}
\end{equation}

The match decision can then be defined as:

\begin{equation} \label{eq:ed2}
\text{Match} =
\begin{cases} 
1, & \text{if } S > 0.7 \\
0, & \text{otherwise}
\end{cases}
\end{equation}

\end{itemize}

After computing all three measures, we used accuracy to report the results obtained from different prompt engineering and parameter settings. Accuracy is defined in equation~\ref{eq:acc}:

\begin{equation} \label{eq:acc}
\text{Accuracy} = \frac{\text{Number of Correct Predictions}}{\text{Total Number of Predictions}}
\end{equation}

\section{Results}
\label{sec:result}

Our evaluation covered different LLMs, focusing on in-context prompt learning, prompt chaining, and temperature settings ranging from 0.2 to 1.0. The experiments were conducted on 24,911 titles and abstracts from the FORC dataset after removing text with empty abstracts. The corresponding results for the candidate LLMs are summarized in Table~\ref{result:llm_prompt}.


Among the tested models, Llama demonstrated the highest overall performance, particularly when utilizing prompt chaining. The results indicated that providing more examples in the prompt — so called few-shot learning — yielded superior outcomes compared to one-shot and zero-shot approaches. For hierarchical classification tasks, prompt chaining, which involves supplying context step-by-step, was especially effective and outperformed standard in-context learning.

The model's performance was also sensitive to temperature settings. While higher temperatures generally improved results, the optimal level was found to be 0.8, after which performance declined. At this temperature, the top model achieved an accuracy of 90.1\% for domain prediction, 80.5\% for subject prediction, and approximately 50\% for topic prediction. Finally, the study found that sentence embedding and string distance were effective evaluation metrics for assessing the models' predictions of research areas.

\begin{table}[!htb]
\centering
\caption{Accuracy of SLMs and LLMs in \% without considering the ORKG taxonomy hierarchy \cite{iceis25}}

\begin{tabular}{|c|c|c|c|c|c|c|}
\hline
\textbf{Model} & \textbf{SLM} & \multicolumn{5}{c|}{\textbf{LLM}} \\
\hline
Temperature &  & \textbf{0.2} & \textbf{0.4} & \textbf{0.6} & \textbf{0.8} & \textbf{1.0} \\
\hline
BERT & 74 & - & - & - & - & - \\
BiLSTM & 66 & - & - & - & - & - \\
Gemma & - & 18 & 24 & 40 & 66 & 62  \\
Llama & - & 34 & 38 & 64 & 82 & 72  \\
Mistral Nemo & - & 14  & 38  & 44 & 76 & 62  \\
Phi & - & 30 & 18 & 44 & 58 & 62 \\
\hline
\end{tabular}
\label{result:slm_prompt}
\end{table}

\vspace{0.5cm}

\begin{table}[!htb]
\centering
\caption{Accuracy of LLMs according to ORKG taxonomy hierarchy in \% across Domain, Subject, and Topic levels using different prompting strategies and evaluation measures. 
EM denotes exact measure, SD is string distance, and ED is embedding distance.}

\begin{tabular}{|p{2.7cm}|ccc|ccc|ccc|}
\hline
\multicolumn{10}{|c|}{\textbf{Zero-shot Prompt}} \\
\hline
\textbf{Hierarchal Level} 
& \multicolumn{3}{c|}{\textbf{Domain}} 
& \multicolumn{3}{c|}{\textbf{Subject}} 
& \multicolumn{3}{c|}{\textbf{Topic}} \\
\hline
\textbf{Model} & \textbf{EM} & \textbf{SD} & \textbf{ED} 
& \textbf{EM} & \textbf{SD} & \textbf{ED} 
& \textbf{EM} & \textbf{SD} & \textbf{ED} \\
\hline
Gemma        & 66.1 & 66.3 & 66.2 & 50.5 & 50.7 & 52.5 & 22.8 & 23.0 & 25.5 \\

Llama        & 83.7 & 83.7 & 83.7 & 54.1 & 54.3 & 56.3 & 27.2 & 27.4 & 30.0 \\
Mistral Nemo & 78.6  &  78.6 &  78.7 & 44.5   & 44.9 &  47.2 &  21.3  &  21.2 & 21.1   \\
Phi          & 57.9  &  57.9 & 57.9  &  33.2 &  33.3 &  33.2 & 17.2 & 17.8 & 18.1  \\ \hline

\end{tabular}

\vspace{0.5cm}

\begin{tabular}{|p{2.7cm}|ccc|ccc|ccc|}
\hline
\multicolumn{10}{|c|}{\textbf{One-shot Prompt}} \\
\hline
\textbf{Hierarchal Level} 
& \multicolumn{3}{c|}{\textbf{Domain}} 
& \multicolumn{3}{c|}{\textbf{Subject}} 
& \multicolumn{3}{c|}{\textbf{Topic}} \\
\hline
\textbf{Model} & \textbf{EM} & \textbf{SD} & \textbf{ED} 
& \textbf{EM} & \textbf{SD} & \textbf{ED} 
& \textbf{EM} & \textbf{SD} & \textbf{ED} \\
\hline
Gemma     & 69.7 & 69.7 & 69.7 & 58.5 & 58.7 & 60.5 & 27.2 & 27.4 & 30.0 \\
Llama        & 85.1  &  85.1 & 85.1  &  57.2 & 57.2  &  59.0 &  30.1 & 30.2  &  32.6 \\
Mistral Nemo & 81.9 & 81.9 & 81.9 & 49.0 & 49.1 & 50.4 & 22.0 &  22.1 & 22.9\\

Phi          & 60.0  &  60.1 & 60.1  &  43.2 &  43.3 &  43.2 & 19.2 & 19.8 & 20.3  \\ \hline
\end{tabular}

\vspace{0.5cm}

\begin{tabular}{|p{2.7cm}|ccc|ccc|ccc|}
\hline
\multicolumn{10}{|c|}{\textbf{Few-shot Prompt}} \\
\hline
\textbf{Hierarchal Level} 
& \multicolumn{3}{c|}{\textbf{Domain}} 
& \multicolumn{3}{c|}{\textbf{Subject}} 
& \multicolumn{3}{c|}{\textbf{Topic}} \\
\hline
\textbf{Model} & \textbf{EM} & \textbf{SD} & \textbf{ED} 
& \textbf{EM} & \textbf{SD} & \textbf{ED} 
& \textbf{EM} & \textbf{SD} & \textbf{ED} \\
\hline
Gemma        & 72.7 & 72.7 & 72.7 & 63.3 & 63.6 & 70.5 & 31.7 & 32.0 & 32.3 \\
Llama        & 87.2  &  87.2 & 87.2  &  71.2 & 71.6  &  72 &  32.8 &  33.2 & 34.0  \\
Mistral Nemo &  83.2 &  83.2 &  83.2 &  52.0 &  52.1 &  52.6 &  23.1 &  23.2 & 23.6   \\
Phi          & 62.2  & 62.2   &  62.2  & 45.6  &  45.8 & 46.0  &  21.0 & 21.2  & 21.5  \\ \hline

\end{tabular}

\vspace{0.5cm}

\begin{tabular}{|p{2.7cm}|ccc|ccc|ccc|}
\hline
\multicolumn{10}{|c|}{\textbf{Prompt Chaining}} \\
\hline
\textbf{Taxonomy Level} 
& \multicolumn{3}{c|}{\textbf{Domain}} 
& \multicolumn{3}{c|}{\textbf{Subject}} 
& \multicolumn{3}{c|}{\textbf{Topic}} \\
\hline
\textbf{Model} & \textbf{EM} & \textbf{SD} & \textbf{ED} 
& \textbf{EM} & \textbf{SD} & \textbf{ED} 
& \textbf{EM} & \textbf{SD} & \textbf{ED} \\
\hline

Gemma        & 89.7 & 89.7 & 89.7 & 74.1 & 74.3 & 76.3 & 37.2 & 37.4 & 40.0 \\
Llama       & 90.1 & 90.1 & 90.1 & 78.5 & 78.7 & 80.5 & 46.8 & 47.0 & 50.5 \\
Mistral Nemo & 89.6  &  89.6 &  89.7 & 74.5   & 74.9 &  77.2 &  25.0 &  26.1 & 36.9  \\
Phi          & 65.6  &  65.6  & 65.6   & 50.1  &  50.1 &  50.6 & 23.2  & 23.4  & 23.9  \\ \hline

\end{tabular}

\label{result:llm_prompt}
\end{table}


\subsection{Error Analysis}


Beyond the automated evaluation described earlier, we carried out a manual error analysis to further examine the performance of the best-performing model. Our focus was on the misclassified outputs generated by Llama 3.1, with the aim of gaining a deeper understanding of its errors.

For this analysis, we randomly selected 100 misclassified scientific texts and manually reviewed the model’s classifications. This process offered qualitative insights into recurring sources of error, such as very short abstracts, missing abstracts, or absent titles, which likely contributed to incorrect predictions. 
Some of publications have short abstracts like with 4-5 words, which might be an issue from data collections.  Some publication do not have abstract, so only title is used for prediction for research areas. Another reason for lower performance of LLMs is that, complex or multiple labels for a science text, such as \textit{"Hydrogeology, Hydrology, Limnology, Urban Water Management, Water Chemistry, Integrated Water Resources Management"}. In this case, LLM predicts research area as Hydrology, so even after applying different evaluation technique, it is hard to consider it a truly predicted.

Overall, the automatic tagging of scientific texts is challenging using LLMs if we go deeper (i.e, 3rd level). One of the key obstacles is the limited availability of cross-domain datasets that adequately represent common subject areas and topics. A limitation of our study stems from the fact that the dataset we employed was originally developed for building a Knowledge Graph and is therefore not an ideal match for the current task in terms of coverage. For instance, certain classes are insufficiently represented: the dataset lacks sub-classes for Arts and Humanities, which restricts its applicability. As a result, our topic classification findings cannot be readily generalized to broader library contexts, where texts span virtually all areas of knowledge.





\section{Conclusion \& Future Work}
\label{sec:future}

This study systematically evaluated the capability of Large Language Models (LLMs) for the automated classification of research areas within scientific literature. Classification was performed against the hierarchical ORKG taxonomy, which includes five top-level domains, 14 subordinate subjects, and 40 further subordinate topics. The findings were derived using advanced prompt-engineering strategies, specifically In-Context Learning and Prompt Chaining. Furthermore, the influence of the LLMs' temperature hyperparameter on classification accuracy was experimentally investigated. Four contemporary Large Language Models — Gemma, Llama, Mistral-Nemo, and Phi — were subjected to a rigorous comparative analysis utilizing the FORC dataset, which comprises 24,911 annotated scientific publications.

Results indicate that LLMs already demonstrate promising single-level classification performance, achieving an accuracy of approximately 80\% when predicting a matching category from all hierarchy levels. When Prompt Chaining was employed, models showed an encouraging ability to correctly classify at each level of the hierarchical taxonomy to a certain extent. However, the models did not achieve satisfactory mastery in the overall multi-level classification task, with accuracy remaining as low as approximately 50\%.


Nevertheless, some limitations emerged. 
A challenge arises from complex or multi-label instances, where a scientific text may be associated with multiple overlapping categories. In such cases, LLMs tend to simplify the prediction to a single domain label (e.g., Hydrology), which — even with alternative evaluation techniques — cannot be regarded as a fully accurate prediction.


Future research should focus on utilizing more advanced LLMs, expanding coverage to finer-grained taxonomy levels, and exploring alternative classification frameworks such as the ACM Computing Classification System and the Dewey Decimal Classification (DDC).  Beyond theoretical contributions, the proposed method can also be deployed in institutional research centers and academic libraries to support systematic identification and categorization of scientific literature.

\section*{Data Sharing} 
We have conducted all experiments on a macro level following
strict data access, storage, and auditing procedures for the sake of
accountability. We release the processed data used in the study along with minimal code to replicate the model for the community. The code and the dataset are available at GitHub.\footnote{\url{https://github.com/Gautamshahi/LLM4ResearchArea}}



\section*{Acknowledgments}
The work has been carried out under the TransforMA project. Authors disclosed receipt of the following financial support for the research, authorship, and/or publication of this article. This project has received funding from the federal-state initiative "Innovative Hochschule" of the Federal Ministry of Education and Research (BMBF) in Germany. 

\bibliographystyle{splncs04}
\bibliography{example}

\begin{thebibliography}{10}
\providecommand{\url}[1]{\texttt{#1}}
\providecommand{\urlprefix}{URL }
\providecommand{\doi}[1]{https://doi.org/#1}

\bibitem{abburi2023generative}
Abburi, H., Suesserman, M., Pudota, N., Veeramani, B., Bowen, E., Bhattacharya, S.: Generative ai text classification using ensemble llm approaches. arXiv preprint arXiv:2309.07755  (2023)

\bibitem{abu2024forc}
Abu~Ahmad, R., Borisova, E., Rehm, G.: Forc@ nslp2024: Overview and insights from the field of research classification shared task. In: International Workshop on Natural Scientific Language Processing and Research Knowledge Graphs. pp. 189--204. Springer (2024)

\bibitem{al2025evaluation}
Al~Nazi, Z., Hossain, M.R., Al~Mamun, F.: Evaluation of open and closed-source llms for low-resource language with zero-shot, few-shot, and chain-of-thought prompting. Natural Language Processing Journal p. 100124 (2025)

\bibitem{auer2019towards}
Auer, S., Mann, S.: Towards an open research knowledge graph. The Serials Librarian  \textbf{76}(1-4),  35--41 (2019)

\bibitem{bird2008acl}
Bird, S., Dale, R., Dorr, B.J., Gibson, B.R., Joseph, M.T., Kan, M.Y., Lee, D., Powley, B., Radev, D.R., Tan, Y.F., et~al.: The acl anthology reference corpus: A reference dataset for bibliographic research in computational linguistics. In: LREC (2008)

\bibitem{bornmann2021growth}
Bornmann, L., Haunschild, R., Mutz, R.: Growth rates of modern science: a latent piecewise growth curve approach to model publication numbers from established and new literature databases. Humanities and Social Sciences Communications  \textbf{8}(1),  1--15 (2021)

\bibitem{chang2024survey}
Chang, Y., Wang, X., Wang, J., Wu, Y., Yang, L., Zhu, K., Chen, H., Yi, X., Wang, C., Wang, Y., et~al.: A survey on evaluation of large language models. ACM transactions on intelligent systems and technology  \textbf{15}(3),  1--45 (2024)

\bibitem{desale2014research}
Desale, S.K., Kumbhar, R.M.: Research on automatic classification of documents in library environment: a literature review. KO KNOWLEDGE ORGANIZATION  \textbf{40}(5),  295--304 (2014)

\bibitem{devatine2024assessing}
Devatine, N., Abraham, L.: Assessing human editing effort on llm-generated texts via compression-based edit distance. arXiv preprint arXiv:2412.17321  (2024)

\bibitem{devlin2018bert}
Devlin, J., Chang, M.W., Lee, K., Toutanova, K.: Bert: Pre-training of deep bidirectional transformers for language understanding. arXiv preprint arXiv:1810.04805  (2018)

\bibitem{enamoto2022multi}
Enamoto, L., Santos, A.R., Maia, R., Weigang, L., Filho, G.P.R.: Multi-label legal text classification with bilstm and attention. International Journal of Computer Applications in Technology  \textbf{68}(4),  369--378 (2022)

\bibitem{feuerriegel2024generative}
Feuerriegel, S., Hartmann, J., Janiesch, C., Zschech, P.: Generative ai. Business \& Information Systems Engineering  \textbf{66}(1),  111--126 (2024)

\bibitem{gao2023prompt}
Gao, A.: Prompt engineering for large language models. Available at SSRN 4504303  (2023)

\bibitem{giglou2024llms4synthesis}
Giglou, H.B., D'Souza, J., Auer, S.: Llms4synthesis: Leveraging large language models for scientific synthesis. arXiv preprint arXiv:2409.18812  (2024)

\bibitem{golub2020automatic}
Golub, K., Hagelb{\"a}ck, J., Ard{\"o}, A.: Automatic classification of swedish metadata using dewey decimal classification: a comparison of approaches. Journal of Data and Information Science  \textbf{5}(1),  18--38 (2020)

\bibitem{hacker2023regulating}
Hacker, P., Engel, A., Mauer, M.: Regulating chatgpt and other large generative ai models. In: Proceedings of the 2023 ACM Conference on Fairness, Accountability, and Transparency. pp. 1112--1123 (2023)

\bibitem{herrmannova2016analysis}
Herrmannova, D., Knoth, P.: An analysis of the microsoft academic graph. D-lib Magazine  \textbf{22}(9/10), ~37 (2016)

\bibitem{hong2021challenges}
Hong, Z., Ward, L., Chard, K., Blaiszik, B., Foster, I.: Challenges and advances in information extraction from scientific literature: a review. JOM  \textbf{73}(11),  3383--3400 (2021)

\bibitem{huang2015bidirectional}
Huang, Z., Xu, W., Yu, K.: Bidirectional lstm-crf models for sequence tagging. arXiv preprint arXiv:1508.01991  (2015)

\bibitem{jaradeh2019open}
Jaradeh, M.Y., Oelen, A., Farfar, K.E., Prinz, M., D'Souza, J., Kismih{\'o}k, G., Stocker, M., Auer, S.: Open research knowledge graph: Next generation infrastructure for semantic scholarly knowledge. In: Proceedings of the 10th international conference on knowledge capture. pp. 243--246 (2019)

\bibitem{jiang2020improving}
Jiang, M., D’Souza, J., Auer, S., Downie, J.S.: Improving scholarly knowledge representation: Evaluating bert-based models for scientific relation classification. In: Digital Libraries at Times of Massive Societal Transition: 22nd International Conference on Asia-Pacific Digital Libraries, ICADL 2020, Kyoto, Japan, November 30--December 1, 2020, Proceedings 22. pp. 3--19. Springer (2020)

\bibitem{kalyan2023survey}
Kalyan, K.S.: A survey of gpt-3 family large language models including chatgpt and gpt-4. Natural Language Processing Journal p. 100048 (2023)

\bibitem{kinney2023semantic}
Kinney, R., Anastasiades, C., Authur, R., Beltagy, I., Bragg, J., Buraczynski, A., Cachola, I., Candra, S., Chandrasekhar, Y., Cohan, A., et~al.: The semantic scholar open data platform. arXiv preprint arXiv:2301.10140  (2023)

\bibitem{liu2024language}
Liu, S., Yu, S., Lin, Z., Pathak, D., Ramanan, D.: Language models as black-box optimizers for vision-language models. In: Proceedings of the IEEE/CVF Conference on Computer Vision and Pattern Recognition. pp. 12687--12697 (2024)

\bibitem{mahapatra2024artificial}
Mahapatra, R., Gayan, M., Jamatia, B., et~al.: Artificial intelligence tools to enhance scholarly communication: An exploration based on a systematic review  (2024)

\bibitem{ollama}
Morgan, J., Chiang, M.: {Ollama}. \url{https://ollama.com} (2024), online; accessed 6 August 2024

\bibitem{mosca2023distinguishing}
Mosca, E., Abdalla, M.H.I., Basso, P., Musumeci, M., Groh, G.: Distinguishing fact from fiction: A benchmark dataset for identifying machine-generated scientific papers in the llm era. In: Proceedings of the 3rd Workshop on Trustworthy Natural Language Processing (TrustNLP 2023). pp. 190--207 (2023)

\bibitem{murphy2022probabilistic}
Murphy, K.P.: Probabilistic machine learning: an introduction. MIT press (2022)

\bibitem{nah2023activity}
Nah, F., Cai, J., Zheng, R., Pang, N.: An activity system-based perspective of generative ai: Challenges and research directions. AIS Transactions on Human-Computer Interaction  \textbf{15}(3),  247--267 (2023)

\bibitem{nsf2023publication}
{National Science Foundation}: Publication output by region, country, or economy, and by scientific field. \url{https://ncses.nsf.gov/pubs/nsb202333/publication-output-by-region-country-or-economy-and-by-scientific-field} (2023), accessed: 2025-09-23

\bibitem{pal2024ai}
Pal, S., Bhattacharya, M., Islam, M.A., Chakraborty, C.: Ai-enabled chatgpt or llm: a new algorithm is required for plagiarism-free scientific writing. International Journal of Surgery  \textbf{110}(2),  1329--1330 (2024)

\bibitem{perelkiewicz2024review}
Pere{\l}kiewicz, M., Po{\'s}wiata, R.: A review of the challenges with massive web-mined corpora used in large language models pre-training. arXiv preprint arXiv:2407.07630  (2024)

\bibitem{pertsas2024annotated}
Pertsas, V., Kasapaki, M., Constantopoulos, P.: An annotated dataset for transformer-based scholarly information extraction and linguistic linked data generation. In: Proceedings of the 9th Workshop on Linked Data in Linguistics@ LREC-COLING 2024. pp. 84--93 (2024)

\bibitem{reimers-2019-sentence-bert}
Reimers, N., Gurevych, I.: Sentence-bert: Sentence embeddings using siamese bert-networks. In: Proceedings of the 2019 Conference on Empirical Methods in Natural Language Processing. Association for Computational Linguistics (11 2019), \url{https://arxiv.org/abs/1908.10084}

\bibitem{rous2012major}
Rous, B.: Major update to acm's computing classification system. Communications of the ACM  \textbf{55}(11),  12--12 (2012)

\bibitem{scott1998dewey}
Scott, M.L.: Dewey decimal classification. Libraries Unlimited  (1998)

\bibitem{iceis25}
Shahi, G., Hummel, O.: On the effectiveness of large language models in automating categorization of scientific texts. In: Proceedings of the 27th International Conference on Enterprise Information Systems - Volume 1: ICEIS. pp. 544--554. INSTICC, SciTePress (2025). \doi{10.5220/0013299100003929}

\bibitem{shahi2024enhancing}
Shahi, G.K., Hummel, O.: Enhancing research information systems with identification of domain experts. In: Proceedings of the Bibliometric-enhanced Information Retrieval Workshop (BIR) at the European Conference on Information Retrieval (ECIR 2024). CEUR Workshop Proceedings, CEUR-WS.org (March 2024)

\bibitem{shahifakecovid}
Shahi, G.K., Nandini, D.: Fake{C}ovid -- a multilingual cross-domain fact check news dataset for covid-19. In: Proceedings of the 14th International {AAAI} {C}onference on {W}eb and {S}ocial {M}edia (2020)

\bibitem{team2024gemma}
Team, G., Mesnard, T., Hardin, C., Dadashi, R., Bhupatiraju, S., Pathak, S., Sifre, L., Rivi{\`e}re, M., Kale, M.S., Love, J., et~al.: Gemma: Open models based on gemini research and technology. arXiv preprint arXiv:2403.08295  (2024)

\bibitem{touvron2023Llama}
Touvron, H., Lavril, T., Izacard, G., Martinet, X., Lachaux, M.A., Lacroix, T., Rozi{\`e}re, B., Goyal, N., Hambro, E., Azhar, F., et~al.: Llama: Open and efficient foundation language models. arXiv preprint arXiv:2302.13971  (2023)

\bibitem{vaswani2017attention}
Vaswani, A., Shazeer, N., Parmar, N., Uszkoreit, J., Jones, L., Gomez, A.N., Kaiser, {\L}., Polosukhin, I.: Attention is all you need. Advances in neural information processing systems  \textbf{30} (2017)

\bibitem{wang2009extensive}
Wang, J.: An extensive study on automated dewey decimal classification. Journal of the American Society for Information Science and Technology  \textbf{60}(11),  2269--2286 (2009)

\bibitem{wang2020microsoft}
Wang, K., Shen, Z., Huang, C., Wu, C.H., Dong, Y., Kanakia, A.: Microsoft academic graph: When experts are not enough. Quantitative Science Studies  \textbf{1}(1),  396--413 (2020)

\bibitem{wang2024gpt}
Wang, S., Hu, T., Xiao, H., Li, Y., Zhang, C., Ning, H., Zhu, R., Li, Z., Ye, X.: Gpt, large language models (llms) and generative artificial intelligence (gai) models in geospatial science: a systematic review. International Journal of Digital Earth  \textbf{17}(1),  2353122 (2024)

\bibitem{young2024chatgpt}
Young, J.S., Lammert, M.: Chatgpt for classification: Evaluation of an automated course mapping method in academic libraries  (2024)

\bibitem{zhang2023usage}
Zhang, C., Tian, L., Chu, H.: Usage frequency and application variety of research methods in library and information science: Continuous investigation from 1991 to 2021. Information Processing \& Management  \textbf{60}(6),  103507 (2023)

\bibitem{zhao2023survey}
Zhao, W.X., Zhou, K., Li, J., Tang, T., Wang, X., Hou, Y., Min, Y., Zhang, B., Zhang, J., Dong, Z., et~al.: A survey of large language models. arXiv preprint arXiv:2303.18223  (2023)

\end{thebibliography}
\end{document}